\documentclass[pra, aps, twocolumn, floatfix, showpacs]{revtex4-1}
\usepackage{graphicx, amsmath, amssymb, times}

\begin{document}
\title{Time-of-flight images of Mott insulators in the Hofstadter-Bose-Hubbard model}
\author{M. Iskin}
\affiliation{
Department of Physics, Ko\c c University, Rumelifeneri Yolu, 34450 Sar{\i}yer, Istanbul, Turkey.
}
\date{\today}
\begin{abstract}

We analyze the momentum distribution function and its artificial-gauge-field 
dependence for the Mott insulator phases of the Hofstadter-Bose-Hubbard model. 
By benchmarking the results of the random-phase approximation (RPA) approach 
against those of the strong-coupling expansion (SCE) for the Landau 
and symmetric gauges, we find pronounced corrections to the former results,
which is a clear manifestation of the critical role played by quantum fluctuations 
in two dimensions.

\end{abstract}
\maketitle

\textit{Introduction:}
The momentum distribution function $n(\mathbf{k})$ of atoms, which is 
defined as the Fourier transform of the one-body density matrix, can be 
directly measured in cold-atom systems by time-of-flight absorption imaging of 
freely expanding gas~\cite{lewenstein07, bloch08, giorgini08}. Since these 
systems are extremely dilute, the atom-atom interactions are negligible during 
such an expansion, and the position of atoms at time $\tau$ are strongly correlated 
with their velocity distribution at the moment of release from the trap, 
i.e., $\mathbf{r} = \hbar \mathbf{k} \tau/m$ with $\hbar$ the Planck constant 
and $m$ the atomic mass. Therefore, the $n(\mathbf{k})$ of atoms has not 
only been the easiest observable to measure but also been routinely used 
for probing distinct phases of matter in atomic systems.

In addition, followed by the recent advances in creating artificial gauge fields 
in atomic systems~\cite{dalibard11, galitski13}, there has been growing 
interest in first the realization of the Hofstadter-type lattice Hamiltonians 
and then the detection of the resultant many-body
phases~\cite{garcia12, struck12, aidelsburger13, miyake13, kennedy15}.
For instance, the MIT group has in their latest preprint measured the 
$n(\mathbf{k})$ of atoms in the superfluid (SF) phase~\cite{kennedy15}, 
revealing both the reduced symmetry of their specific gauge field and the 
resultant degeneracy of the ground state~\cite{leblanc15}. There is no doubt that 
such a capacity to tune strong gauge fields together with strong interactions 
paves ultimately the way for creating and observing uncharted many-body 
phases and transitions in between, one of the immediate candidates of 
which is the renowned SF-MI transition~\cite{lewenstein07, bloch08}.

Motivated by these recent works, in this brief paper, we study $n(\mathbf{k})$ 
of atoms for the MI phases of the Hofstadter-Bose-Hubbard 
model on a square lattice. For this purpose, we compare the results of 
RPA and SCE approaches for the Landau and symmetric gauges, and 
find substantial corrections to the former results depending strongly on 
the specified gauge.

\textit{Hamiltonian and Phase Diagram:}
These results are obtained for the following Hamiltonian
\begin{align}
H = -\sum_{ij} t_{ij} c_i^\dagger c_j +
 \frac{U}{2} \sum_i \widehat{n}_i (\widehat{n}_i-1) - \mu \sum_i \widehat{n}_i,
\label{eqn:ham} 
\end{align}
where the hopping parameter $t_{ij} = t e^{i \theta_{ij}}$ connects 
nearest-neighbor sites with phase factor $\theta_{ij}$ taking the gauge fields 
into account, $c_i^\dagger$ ($c_i$) creates (annihilates) a boson on site $i$,
the boson-boson interaction is on-site and repulsive $U \ge 0$,
$\widehat{n}_i = c_i^\dagger c_i$ is the number operator, 
and $\mu \ge 0$ is the chemical potential.
In this paper, we compare the results of the usual
$(A)$ no-gauge limit, where $\theta_{ij} = 0$ for all hoppings;
with those of
$(B)$ Landau gauge, where $\theta_{ij} = 2\pi\phi u$ for $(u,v)$ to $(u,v+1)$ 
and $0$ for $(u,v)$ to $(u+1,v)$ hoppings;
$(C)$ symmetric gauge, where $\theta_{ij} = \pi\phi u$ for $(u,v)$ to $(u,v+1)$ 
and $-\pi\phi v$ for $(u,v)$ to $(u+1,v)$ hoppings;
and $(D)$ MIT gauge~\cite{kennedy15}, where $\theta_{ij} = 2\pi\phi (u+v) $ 
for $(u,v)$ to $(u,v+1)$ and $0$ for $(u,v)$ to $(u+1,v)$ hoppings.
Here, $(u,v)$ corresponds to the Cartesian coordinates of site $i$, 
and $\theta_{ij}$ are chosen such that the magnetic flux $\phi = p/q$ is 
the same for all gauges, where $p$ and $q$ are co-prime numbers 
with $p \le q$.

\begin{figure} [htb]
\centerline{\scalebox{0.21}{\includegraphics{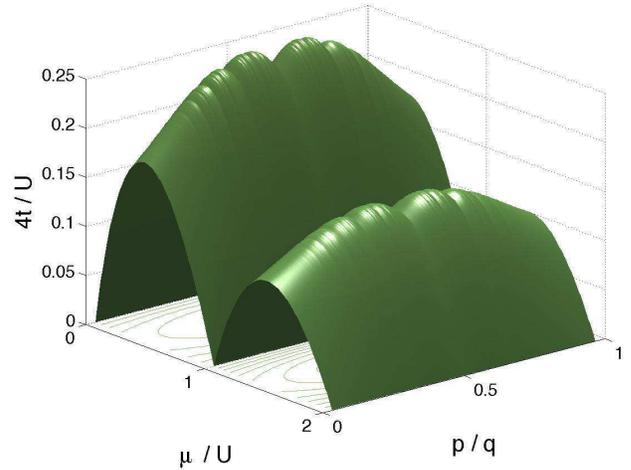}}}
\caption{\label{fig:pd} (Color online)
The ground-state phase diagram is shown as a function of chemical potential 
$\mu$, magnetic flux $\phi = p/q$ and hopping strength $4t$.
}
\end{figure}

In the atomic ($t = 0$) limit, since $H$ commutes with $\widehat{n}_i$, the
thermal average $n_i = \langle \widehat{n}_i \rangle$ is such that the 
ground-state energy is minimised for a given $\mu$, leading to a uniform 
occupation ($n_i = n$) of bosons thanks to the translational invariance of $H$. 
When $U = 0$ and $\mu = 0$, the spectrum of $H$ corresponds to the 
celebrated Hofstadter butterfly~\cite{hofstadter76, kohmoto89}. 
It is also very well-known that the range of $\mu$ about which the ground state 
is a MI with an integer occupation $n$ decreases as a function of increasing 
$t/U$, and depending on $n$ and $\phi$, the MIs disappear at a critical value 
of $t/U$, beyond which the system becomes a SF~\cite{niemeyer99}. 
For instance, the qualitative phase diagram of $H$ can be obtained within the 
mean-field approximation, e.g., the decoupling or variational Gutzwiller 
techniques, leading to~\cite{umucalilar07, goldbaum09, iskina}
\begin{align}
\label{eqn:mf}
\frac{1}{\epsilon^{pq}} = \frac{n+1}{U n - \mu} - \frac{n}{U (n-1) - \mu}
\end{align}
at zero temperature for the MI-SF phase transition boundary, where $n \ge 0$ 
is an integer number. Here, $\epsilon^{pq}$ is the minimal eigenvalue 
of the hopping matrix $\sum_j (-t_{ij}) f_{j} = \epsilon^{pq} f_i$ and it corresponds
to the maximal single-particle kinetic energy of the Hofstadter butterfly,
e.g., $\epsilon^{0} = 4t$ when $\phi = 0$.
Since the effects of $\theta_{ij}$ enter Eq.~(\ref{eqn:mf}) through its dependence
on $\epsilon^{pq}$, the mean-field phase boundary is clearly independent of 
the gauge, which is simply because only the position in the magnetic Brillouin 
zone but not the value of $\epsilon^{pq}$ depends on the gauge. 
However, this is not the case for the SF properties which are gauge 
dependent within the mean-field approaches. 

In Fig.~\ref{fig:pd}, we show the ground-state phase diagram as a function of $\mu$, 
$\phi = p/q$ and $4t$, which is obtained by solving Eq.~(\ref{eqn:mf}) together with
the Harper's equation. 
Both the symmetry around $p/q = 1/2$ and the intriguing 
structure of the MI-SF phase transition boundary are due to the dependence of 
$\epsilon^{pq}$ on $\phi$~\cite{hofstadter76, kohmoto89}.
In addition, the incompressible (compressible) MI (SF) phase grows (shrinks) 
when $\phi$ increases from $0$, a consequence of which is due to the localizing 
effects of magnetic flux on particles, and all of these results are in agreement 
with earlier findings~\cite{niemeyer99, umucalilar07, goldbaum09, iskina}. 
Having introduced the model Hamiltonian and reviewed its phase diagram,
next we are ready to discuss the momentum distribution of bosons 
for the MIs.

\textit{Momentum Distribution:}
As discussed in the \textit{Introduction}, the $n(\mathbf{k})$ of atoms corresponds
to the Fourier transform of the one-body density matrix, and it is given 
by~\cite{freericks09, iskin09, moller10, sinha11} 
\begin{align}
\label{eqn:nk}
n(\mathbf{k}) = \frac{|w(\mathbf{k})|^2}{M} 
\sum_{jj'} \langle c_{j'}^\dagger c_j \rangle e^{i\mathbf{k} \cdot (\mathbf{r_{j'}} - \mathbf{r_j})},
\end{align}
where $M$ is the number of sites and $\mathbf{r_j} = (ua, va)$ is the position 
of site $j$ with $a$ the lattice spacing. In the following, we set 
the Fourier transform of the Wannier function $w(\mathbf{k})$ to $1$, since
it depends on the particular optical lattice potential and has nothing to do 
with our $H$.

In this paper, we calculate $n(\mathbf{k})$ for the MIs using two approaches: 
$(I)$ RPA~\cite{sinha11, iskin09} and 
$(II)$ SCE in $t/U$~\cite{freericks09, iskin09}. 
We emphasize that while the result of the RPA approach corresponds to the exact 
$n(\mathbf{k})$ only in the limit of infinite dimensions and zero magnetic flux, 
the results of the SCE approach are exact in two dimensions for the specified 
gauges up to the given order in $t/U$.

\textit{(I) Random-Phase Approximation:}
In the RPA approach~\cite{sinha11, iskin09}, since the thermal averages of products 
of operators are replaced by the product of their thermal averages, the fluctuations 
are not fully taken into account. After a lengthy but straightforward algebra, 
one finds
\begin{align}
\label{eqn:RPA}
n_\textrm{RPA}^{pq} (\mathbf{k}) = \frac{1}{2q} \sum_{\ell = 0}^{q-1} 
\frac{\varepsilon_\ell^{pq} (\mathbf{k}) + \widetilde{U}} 
{\sqrt{\left[\varepsilon_\ell^{pq} (\mathbf{k})\right]^2 + 2\widetilde{U} \varepsilon_\ell^{pq} (\mathbf{k}) + U^2}}
- \frac{1}{2}
\end{align}
for a MI with $n$ bosons per site at zero temperature, where 
$\widetilde{U} = U(2n+1)$ and $\varepsilon_\ell^{pq} (\mathbf{k})$ is the energy 
dispersion of a single particle in the $\ell$th Hofstadter band.
Note that the form of Eq.~(\ref{eqn:RPA}) is exactly the same as the 
usual Bose-Hubbard model, \textit{i.e.}, the main difference is a sum over 
the Hofstadter bands, and that it has an overall factor of 
$1/q$ in comparison to the one given in Ref.~\cite{sinha11}.
While the set of $\varepsilon_\ell^{pq} (\mathbf{k})$ values depends only 
on $\phi$ and lattice geometry, their corresponding positions in the 1st magnetic 
Brillouin zone, and therefore $n(\mathbf{k})$, are gauge 
dependent~\cite{moller10, sinha11}.
For instance, $n(\mathbf{k})$ exhibits $q$ peaks as a function of $\mathbf{k}$, 
and only the number $q$ but not the positions are controlled by $\phi$. 
Note that $\epsilon^{pq} \equiv \max\{ \varepsilon_\ell^{pq} (\mathbf{k}) \}$ in 
Eq.~(\ref{eqn:mf}) which is also a gauge-independent quantity as remarked above.
In particular, when $\phi = 0$, a $d$-dimensional hypercubic lattice gives
rise to a single band with dispersion 
$\varepsilon^0 (\mathbf{k}) = -2t \sum_{k_i, i =1}^d \cos(k_i a)$, 
and it is already established that $n_\textrm{RPA}^0 (\mathbf{k})$ becomes 
exact as $d \to \infty$ while keeping $d t$ fixed~\cite{freericks09, iskin09}. 

To compare Eq.~(\ref{eqn:RPA}) with our exact results of the SCE approach 
derived below, let us expand $n_\textrm{RPA}^{pq} (\mathbf{k})$ in a 
power series up to $3$rd order in $t/U$, leading to
\begin{align}
\label{eqn:rpank}
n_\textrm{RPA}^{pq} (\mathbf{k}) & =  
n 
- \frac{2n (n+1)}{q U} \sum_{\ell = 0}^{q-1} \varepsilon_\ell^{pq} (\mathbf{k}) \nonumber \\
& + \frac{3n (n+1) (2n+1)}{q U^2} \sum_{\ell = 0}^{q-1} \left[ \varepsilon_\ell^{pq} (\mathbf{k}) \right]^2 \nonumber \\
&- \frac{4n (n+1) (5n^2+5n+1)}{q U^3} \sum_{\ell = 0}^{q-1} \left[ \varepsilon_\ell^{pq} (\mathbf{k}) \right]^3.
\end{align}
For a given $\phi$, the sums over Hofstadter bands can be easily evaluated 
for a given gauge by noting
$
\sum_{\ell = 0}^{q-1} \left[\varepsilon_\ell^{pq} (\mathbf{k})\right]^s 
= \textrm{Trace} \big \lbrace \left[T^{pq}(\mathbf{k})\right]^s \big\rbrace,
$
where $T^{pq}(\mathbf{k})$ describes the kinetic energy of a single particle 
in the $1$st magnetic Brillouin zone. 

For instance, $T^{pq}(\mathbf{k})$ is a $q \times q$ matrix in the Landau 
gauge~\cite{hofstadter76, kohmoto89}
%
%%%
\begin{equation}
\label{eqn:}
\begin{bmatrix}
h_0 & -te^{ik_x a} & 0 & . & -te^{-ik_x a} \\
-te^{-ik_x a} & h_1  & -te^{ik_x a} & . & 0 \\
0 & -te^{-ik_x a} & h_2 & . & . \\
. & . & . & . & -te^{ik_x a} \\
-te^{ik_x a}  & 0 & . & -te^{-ik_x a} & h_{q-1}
\end{bmatrix}
\end{equation}
with $h_\ell = -2t \cos(k_y a + 2\pi \phi \ell)$, for which the $s = 1$ trace
equals to $-2t [\cos(k_xa) + \cos(k_ya)]$ when $(p,q) = (1,1)$ 
and it vanishes for $q > 1$; the $s = 2$ trace
equals to $4t^2 \left[\cos(k_xa) + \cos(k_ya)\right]^2$ when $(p,q) = (1,1)$, 
to $8t^2\left[\cos^2(k_xa) + \cos^2(k_ya)\right]$ when $(p,q) = (1,2)$ 
and to $4qt^2$ for $q > 2$; and lastly the $s = 3$ trace
equals to $-8t^3 \left[\cos(k_xa) + \cos(k_ya)\right]^3$ when $(p,q) = (1,1)$
and to $-6t^3\left[\cos(3k_xa) + \cos(3k_ya)\right]$ when $q = 3$, 
but it vanishes for $q > 3$. Thus, Eq.~(\ref{eqn:rpank}) reduces to
\begin{widetext}
\begin{align}
\label{eqn:RPA11}
n_\textrm{RPA}^{11} (\mathbf{k}) & = 
n + \frac{4n (n+1)}{(U/t)} [\cos(k_x a) + \cos(k_y a)]
 + \frac{12n (n+1) (2n+1)}{(U/t)^2} [\cos(k_x a) + \cos(k_y a)]^2 \nonumber  \\
& + \frac{32n (n+1) (5n^2+5n+1)}{(U/t)^3} [\cos(k_x a) + \cos(k_y a)]^3, \\
\label{eqn:RPA12}
n_\textrm{RPA}^{12} (\mathbf{k}) & = 
n + \frac{6n (n+1) (2n+1)}{(U/t)^2} [\cos(2k_x a) + \cos(2k_y a) + 2] + \mathcal{O} (t/U)^4, \\
\label{eqn:RPAp3}
n_\textrm{RPA}^{p3} (\mathbf{k}) & = 
n + \frac{12n (n+1) (2n+1)}{(U/t)^2} + \frac{8n (n+1) (5n^2+5n+1)}{(U/t)^3} [\cos(3k_x a) + \cos(3k_y a)], \\
\label{eqn:RPApq}
n_\textrm{RPA}^{p,q > 3} (\mathbf{k}) & = n + \frac{12n (n+1) (2n+1)}{(U/t)^2} + \mathcal{O} (t/U)^4.
\end{align}
\end{widetext}
Equations~(\ref{eqn:RPA11}-\ref{eqn:RPApq}) clearly show that 
the first $\mathbf{k}$ dependence of $n_\textrm{RPA}^{pq} (\mathbf{k})$ arises 
at the $q$th order in $t/U$. 
More importantly, we note that Eqs.~(\ref{eqn:RPA11}-\ref{eqn:RPApq}) are 
symmetric in $k_x$ and $k_y$ even though the spatial symmetry between 
$x$ and $y$ directions is explicitly broken by the Landau gauge.
Note also that Eq.~(\ref{eqn:RPA11}) coincides with that of the $\phi = 0$ 
result since $\varepsilon_\ell^{pq} (\mathbf{k})$ is a periodic function 
of $\phi$ with a period of $1$~\cite{hofstadter76, kohmoto89}.
Unlike the $\phi = 0$ case for which the RPA approach captures the essential 
features of $n^{0} (\mathbf{k})$ even in finite dimensions~\cite{freericks09, iskin09}, 
next we use the SCE approach and show that the corrections to 
$n_\textrm{RPA}^{pq} (\mathbf{k})$ are quite dramatic in the presence of 
gauge fields in two dimensions.

\textit{(II) Strong-Coupling Expansion:}
In the SCE approach~\cite{freericks09, iskin09}, the wave function of MIs 
is achieved via a many-body perturbation theory in the kinetic energy term 
up to $3$rd order in $t/U$. 
In principle, one can apply the perturbation theory on the $0$th-order wave 
function 
$
|\Psi_{\rm MI}^{(0)} \rangle = \prod_{j = 1}^{M} 
\left( c_j^\dagger \right)^{n} | 0 \rangle / \sqrt{n!},
$
where $| 0 \rangle$ is the vacuum state,
and calculate $|\Psi_{\rm MI} \rangle$ up to any desired order. However, since 
the number of intermediate states increases dramatically, here we perform 
this expansion only up to $3$rd order in $t/U$, and obtain
$
|\Psi_{\rm MI} \rangle = |\psi_{\rm MI} \rangle/ A
$
where
\begin{align}
\label{eqn:wf-non}
|\psi_{\rm MI} \rangle &=
| \Psi_{\rm MI}^{(0)} \rangle 
+ \sum_{m'} \frac{V_{m' 0}}{E_{0m'}} | m' \rangle
+ \sum_{m'm''} \frac{V_{m''m'} V_{m'0}}{E_{0m''} E_{0m'}} | m'' \rangle \nonumber \\
& + \sum_{m'm''m'''} \frac{V_{m'''m''} V_{m''m'} V_{m'0}}{E_{0m'''} E_{0m''} E_{0m'}} | m''' \rangle
+ \cdots
\end{align}
is the unnormalized wave function which needs to be divided by a proper 
normalization coefficient $A$ in order to get the correct order of perturbation.
Here,
$
V_{m'0} = -\sum_{jj'} t_{jj'} \langle m' | c_j^\dagger c_{j'} | \Psi_{\rm MI}^{(0)} \rangle
$ 
connects the $1$st-order intermediate states $| m' \rangle$ to 
$| \Psi_{\rm MI}^{(0)} \rangle$, $E_{0m'} = E_{\rm MI}^{(0)}-E_{m'}^{(0)}$ is 
their $0$th-order energy difference, and $| m'' \rangle$ and $| m''' \rangle$
are respectively the $2$nd and $3$rd-order intermediate states.
Note that while $| \Psi_{\rm MI}^{(0)} \rangle$ and $| m' \rangle$,
$| m' \rangle$ and $| m'' \rangle$, and $| m'' \rangle$ and $| m''' \rangle$ 
states are connected to each other with a single hopping,
$| m'' \rangle$ and $| m''' \rangle$ states must be different from
the $| \Psi_{\rm MI}^{(0)} \rangle$ state. Therefore, the normalization 
condition $\langle \Psi_{\rm MI} | \Psi_{\rm MI} \rangle = 1$ gives 
$
A^2 = 1 + 4 n (n+1) M t^2/U^2 + O(t/U)^4,
$
which has vanishing $1$st and $3$rd order terms.

After a very lengthy and tedious algebra, one finds
\begin{align}
\label{eqn:obc}
\langle \Psi_{\rm MI} | a_{j'}^\dagger & a_j | \Psi_{\rm MI} \rangle  = 
n \delta_{jj'} 
 + \frac{2n (n+1)}{U} t_{jj'} \nonumber \\
& + \frac{3n (n+1) (2n+1)}{U^2} \left(\sum_{j_1} t_{jj_1} t_{j_1j'} - 4t^2 \delta_{jj'} \right) \nonumber \\
& + \frac{4n (n+1) (5n^2+5n+1)}{U^3} \sum_{j_1 j_2} t_{jj_2} t_{j_2j_1} t_{j_1j'} \nonumber \\
& - \frac{n (n+1) (131n^2+131n+26)}{U^3} t^2 t_{jj'}
\end{align}
for a square lattice with nearest-neighbor hopping at zero temperature. 
We note in Eq.~(\ref{eqn:obc}) that the $2$ terms that are explicitly proportional 
to $t^2$ are finite-$d$ corrections, including the $2$nd term in the $2$nd line
and the $4$th line, as they vanish in the $d \to \infty$ limit while 
keeping $d t$ fixed. Since Eq.~(\ref{eqn:obc}) is derived exactly using a
generic hopping matrix $t_{ij}$, we are ready to benchmark it against 
the results of the RPA approach for a number of specified gauges.
For this purpose, we make use of the following identities: the sum
$
\sum_{\ell = 0}^{q-1} \cos(\alpha - 2 n \pi \phi \ell)
$
equals to $q \cos(\alpha)$ when $q = n$ and it vanishes for $q > n$; 
the sum
$
\sum_{\ell = 0}^{q-1} \cos^2(\alpha - 2\pi \phi \ell)
$
equals to $q\cos^2(\alpha)$ when $(p,q) = \{ (1,1), (1,2) \}$ and $q/2$ for $q > 2$;
and the sum
$
\sum_{\ell = 0}^{q-1} \cos^3(\alpha - 2\pi \phi \ell)
$
equals to $\cos^3(\alpha)$ when $(p,q) = (1,1)$ and to 
$3 \cos(3\alpha)/4$ when $q = 3$, but it vanishes for $(p,q) = (1,2)$ or $q > 3$.

\textit{(II-A) No-Gauge Limit:}
Setting $\theta_{ij} = 0$ for all hoppings in Eq.~(\ref{eqn:obc}), we obtain
\begin{align}
\label{eqn:n0}
n^0 (\mathbf{k}) & =  
n - \frac{2n (n+1)}{U} \varepsilon^0(\mathbf{k}) \nonumber \\
& + \frac{3n (n+1) (2n+1)}{U^2} \{[\varepsilon^0 (\mathbf{k})]^2 - 4 t^2\} \nonumber \\
& - \frac{4n (n+1) (5n^2+5n+1)}{U^3} [\varepsilon^0(\mathbf{k})]^3 \nonumber \\
& + \frac{n (n+1) (131n^2+131n+26)}{U^3} t^2 \varepsilon^0(\mathbf{k}), 
\end{align}
where $\varepsilon^0 (\mathbf{k}) = -2t \left[ \cos(k_x a) + \cos(k_y a) \right]$ 
is the usual dispersion relation for a square lattice.
Since the two terms that are explicitly proportional to $t^2$ are finite-$d$ 
corrections, they are not captured by the result of the RPA approach that 
is given in Eq.~(\ref{eqn:RPA11}).

\textit{(II-B) Landau Gauge:}
On the other hand, setting $\theta_{ij} = 2\pi\phi u$ for $(u,v)$ to $(u,v+1)$ 
and $0$ for $(u,v)$ to $(u+1,v)$ hoppings in Eq.~(\ref{eqn:obc}), we obtain
\begin{widetext}
\begin{align}
\label{eqn:L11}
n_\textrm{L}^{11} (\mathbf{k}) & =  
n + \frac{4n (n+1)}{(U/t)} [\cos(k_x a) + \cos(k_y a)]
 + \frac{12n (n+1) (2n+1)}{(U/t)^2} \big\lbrace [\cos(k_x a) + \cos(k_y a)]^2 - 1 \big\rbrace \\
& + \frac{32n (n+1) (5n^2+5n+1)}{(U/t)^3} [\cos(k_x a) + \cos(k_y a)]^3 
 - \frac{2n (n+1) (131n^2+131n+26)}{(U/t)^3} [\cos(k_x a) + \cos(k_y a)], \nonumber \\
\label{eqn:L12}
n_\textrm{L}^{12} (\mathbf{k}) & = 
n + \frac{4n (n+1)}{(U/t)} \cos(k_x a) 
 + \frac{6n (n+1) (2n+1)}{(U/t)^2} [\cos(2k_x a) + \cos(2k_y a)] \\
& + \frac{32n (n+1) (5n^2+5n+1)}{(U/t)^3} \cos(k_x a) [\cos^2(k_x a) + \cos^2(k_y a)] 
 - \frac{2n (n+1) (131n^2+131n+26)}{(U/t)^3} \cos(k_x a), \nonumber \\
\label{eqn:Lp3}
n_\textrm{L}^{p3} (\mathbf{k}) & = 
n + \frac{4n (n+1)}{(U/t)} \cos(k_x a)
 + \frac{6n (n+1) (2n+1)}{(U/t)^2} \cos(2k_x a) \\
& + \frac{8n (n+1) (5n^2+5n+1)}{(U/t)^3} [\cos(3k_x a) + \cos(3k_y a) + 6\cos(k_x a)]
 - \frac{2n (n+1) (131n^2+131n+26)}{(U/t)^3} \cos(k_x a), \nonumber \\
\label{eqn:Lpq}
n_\textrm{L}^{p,q > 3} (\mathbf{k}) & = 
n + \frac{4n (n+1)}{(U/t)} \cos(k_x a)
 + \frac{6n (n+1) (2n+1)}{(U/t)^2} \cos(2k_x a) \\
& + \frac{8n (n+1) (5n^2+5n+1)}{(U/t)^3} \big\lbrace \cos(3k_x a) + [7 + 2\cos(2\pi p/q)] \cos(k_x a) \big\rbrace
 - \frac{2n (n+1) (131n^2+131n+26)}{(U/t)^3} \cos(k_x a). \nonumber
\end{align}
\end{widetext}
Note that Eq.~(\ref{eqn:L11}) exactly coincides with 
Eq.~(\ref{eqn:n0}) since $\phi = 1$ and $0$ are equivalent in this gauge.
We also note that, unlike the results of the RPA approach that are given in 
Eqs.~(\ref{eqn:RPA11}-\ref{eqn:RPApq}), these exact results are not symmetric 
in $k_x$ and $k_y$, showing that it is only the first $k_y$ dependence that 
arises at the $q$th order in $t/U$. This is not surprising because while 
the one-body correlation operator $c_{j'}^\dagger c_j$ connects 
$|\Psi_{\rm MI} \rangle$ to itself at the $1$st order in $x$ direction, 
the connection is established at the $q$th order in $y$ direction due to 
the presence of $2\pi\phi u$. In addition, on top of the RPA 
contributions, Eqs.~(\ref{eqn:L11}-\ref{eqn:Lpq}) contain various other terms, 
showing that the finite-$d$ corrections are quite substantial in the presence of 
gauge fields in two dimensions~\cite{footnote}. Thus, one of our main conclusions 
in this paper is that the mismatch between the results of RPA and SCE 
approaches grows so dramatically as $q$ increases from $1$ that the former 
approach fails to reproduce any of the exact terms up to $3$rd order in $t/U$ 
for $q > 3$.

\textit{(II-C) Symmetric Gauge:}
Similarly, setting $\theta_{ij} = \pi\phi u$ for $(u,v)$ to $(u,v+1)$ and $-\pi\phi v$ 
for $(u,v)$ to $(u+1,v)$ hoppings in Eq.~(\ref{eqn:obc}), we obtain
\begin{align}
\label{eqn:S11}
n_\textrm{S}^{11} (\mathbf{k}) &=  
n + \frac{12n (n+1) (2n+1)}{(U/t)^2} \big\lbrace [\cos(k_x a) \nonumber \\ 
& + \cos(k_y a)]^2 - 1 \big\rbrace + \mathcal{O}(t/U)^4, \\
\label{eqn:Spq}
n_\textrm{S}^{p,q > 1} (\mathbf{k}) & =  n + \mathcal{O} (t/U)^4.
\end{align}
Note that Eq.~(\ref{eqn:S11}) does not reproduce 
Eq.~(\ref{eqn:n0}) since $\phi = 1$ and $0$ are not equivalent in this gauge. 
We also note that, unlike the results of the SCE approach for the Landau 
gauge that are given in Eqs.~(\ref{eqn:L11}-\ref{eqn:Lpq}), here 
the $\mathbf{k}$ dependence is not only symmetric in $k_x$ and $k_y$, 
thanks to the spatial symmetry between $x$ and $y$ directions, but also 
the first $\mathbf{k}$ dependence arises at the $2q$th order in $t/U$. 
This is also not surprising because the one-body correlation operator 
$c_{j'}^\dagger c_j$ connects $|\Psi_{\rm MI} \rangle$ to itself at the
$2q$th order in both directions due to the presence of $\pi\phi u$.
In addition, the $\mathbf{k}$-independent $2$nd order term in Eq.~(18) 
is a finite-$d$ correction to the result of the RPA approach in this gauge. 
Therefore, $n_S^{pq}(\mathbf{k})$ becomes more and more featureless function 
of $\mathbf{k}$ as $q$ increases from $1$, especially deep in the 
MIs when $t/U$ is very small.

\textit{(II-D) MIT Gauge:}
Lastly, setting $\theta_{ij} = 2\pi\phi (u+v) $ for $(u,v)$ to $(u,v+1)$ 
and $0$ for $(u,v)$ to $(u+1,v)$ hoppings in Eq.~(\ref{eqn:obc}) leads 
exactly to Eqs.~(\ref{eqn:L11}-\ref{eqn:Lpq}), and therefore, 
the MIT~\cite{kennedy15} and Landau gauges have exactly the 
same $n(\mathbf{k})$.

\textit{Conclusions:} 
To summarize, we studied the expansion images of atoms for the MI 
phases of the Hofstadter-Bose-Hubbard model on a square lattice. 
In particular, we explicitly calculated the momentum distribution function 
for the Landau and symmetric gauges with both RPA and SCE approaches, 
and found marked corrections to the former results depending strongly 
on the specified gauge. Such a comparison clearly manifests the importance
of the critical role played by quantum fluctuations in two dimensions.

\textit{Acknowledgments:} 
We gratefully acknowledge funding from T\"{U}B$\dot{\mathrm{I}}$TAK Grant No. 1001-114F232.


\begin{thebibliography}{99}

% cold atom reviews
\bibitem{lewenstein07} M. Lewenstein, A. Sanpera, V. Ahufinger, B. Damski, A. Sen De, and U. Sen, Adv. Phy. \textbf{56}, 243 (2007).
\bibitem{bloch08} I. Bloch, J. Dalibard, and W. Zwerger, Rev. Mod. Phys. \textbf{80}, 885 (2008).
\bibitem{giorgini08} S. Giorgini, L. P. Pitaevskii, and S. Stringari, Rev. Mod. Phys. \textbf{80}, 1215 (2008).

% artificial gauge fields review
\bibitem{dalibard11} J. Dalibard, F. Gerbier, G. Juzelinas, and P. \"{O}hberg, Rev. Mod. Phys. {\bf 83}, 1523 (2011).
\bibitem{galitski13} V. Galitski and I. B. Spielman, Nature {\bf 494}, 49 (2013).

% Hofstadter exps
\bibitem{garcia12}  K. Jim\'enez-Garc\'ia, L. J. LeBlanc, R. A. Williams, M. C. Beeler, A. R. Perry, and I. B. Spielman, Phys. Rev. Lett. \textbf{108}, 225303 (2012). % peierls
\bibitem{struck12}  J. Struck, C. \"Olschl\"ager, M. Weinberg, P. Hauke, J. Simonet, A. Eckardt, M. Lewenstein, K. Sengstock, and P. Windpassinger, Phys. Rev. Lett. \textbf{108}, 225304 (2012). % driven ol - Peierls
\bibitem{aidelsburger13} M. Aidelsburger, M. Atala, M. Lohse, J. T. Barreiro, B. Paredes, and I. Bloch, Phys. Rev. Lett. \textbf{111}, 185301 (2013). % realization of hofstadter ham
 \bibitem{miyake13} H. Miyake, G. A. Siviloglou, C. J. Kennedy, W. C. Burton, and W Ketterle, Phys. Rev. Lett. \textbf{111}, 185302 (2013). % realizing Harper ham - QH
\bibitem{kennedy15} C. J. Kennedy, W. C. Burton, W. C. Chung, and W. Ketterle, arXiv:1503.08243 (2015). %HBH realization

\bibitem{leblanc15}  See also: L. J. LeBlanc, K. Jim\'enez-Garc�a, R. A. Williams, M. C. Beeler, W. D. Phillips, and I. B. Spielman,  arXiv:1502.07443 (2015) for a somewhat related work in continuum. %gauge matters

% Hofstadter butterfly
\bibitem{hofstadter76} D. R. Hofstadter, Phys. Rev. B \textbf{14}, 2239 (1976).
\bibitem{kohmoto89} M. Kohmoto, Phys. Rev. B \textbf{39}, 11943 (1989).

% MI-SF transition
\bibitem{niemeyer99} M. Niemeyer, J. K. Freericks, and H. Monien, Phys. Rev. B \textbf{60}, 2357 (1999).
\bibitem{umucalilar07} R. O. Umucal{\i}lar and M. \"O. Oktel, Phys. Rev. A \textbf{76}, 055601 (2007).
\bibitem{goldbaum09} D. S. Goldbaum and E. J. Mueller, Phys. Rev. A \textbf{79}, 021602(R) (2009).
\bibitem{iskina} M. Iskin, Eur. Phys. J. B \textbf{85}, 76 (2012).
 
% mom. dist.
\bibitem{freericks09} J. K. Freericks, H. R. Krishnamurthy, Y. Kato, N. Kawashima, and N. Trivedi, Phys. Rev. A \textbf{79}, 053631 (2009).
\bibitem{iskin09} M. Iskin and J. K. Freericks, Phys. Rev. A \textbf{80}, 063610 (2009).
\bibitem{sinha11} S. Sinha and K. Sengupta, EPL \textbf{93}, 30005 (2011). %RPA
\bibitem{moller10} G. M\"oller and N. R. Cooper, Phys. Rev. A \textbf{82}, 063625 (2010). 

\bibitem{footnote} We note that since the $\mathbf{k}$-independent $2$nd order 
RPA terms that are found in Eqs.~(\ref{eqn:RPAp3}-\ref{eqn:RPApq}) are 
coincidentally canceled by the finite-$d$ correction that is found in 
Eq.~(\ref{eqn:obc}), these terms do not appear in Eqs.~(\ref{eqn:Lp3}-\ref{eqn:Lpq}).

\end{thebibliography}
\end{document}